\documentclass[12pt]{article}
\usepackage{amsmath,amssymb,epsfig}
\makeatletter \@addtoreset{equation}{section} \makeatother
\renewcommand{\theequation}{\thesection.\arabic{equation}}
\newcommand{\ba}{\begin{array}}
\newcommand{\ea}{\end{array}}
\newcommand{\beq}{\begin{equation}}
\newcommand{\eeq}{\end{equation}}
\newcommand{\bea}{\begin{eqnarray}}
\newcommand{\eea}{\end{eqnarray}}
\def\bce{\begin{center}}
\def\ece{\end{center}}

\def\nonu{\nonumber}

\def\pa{\partial}
\def\al{\alpha}
\def\be{\beta}

\def\de{\delta}

\def\la{\lambda}

\def\diag{\mathop{\rm diag}}

\def\eps6{{\displaystyle \mathop{\epsilon}^{6}}{}}

\def\nab6{{\displaystyle \mathop{\nabla}^{6}}{}}
\def\0{{\sst{(0)}}}
\def\1{{\sst{(1)}}}
\def\2{{\sst{(2)}}}
\def\3{{\sst{(3)}}}
\def\4{{\sst{(4)}}}
\def\5{{\sst{(5)}}}
\def\6{{\sst{(6)}}}
\def\7{{\sst{(7)}}}
\def\8{{\sst{(8)}}}

\def\ba{\begin{array}}
\def\ea{\end{array}}
\def\beq{\begin{equation}}
\def\eeq{\end{equation}}
\def\be{\begin{equation}}
\def\ee{\end{equation}}

\def\Tr{\mathop{\rm Tr}}

\def\diag{\mathop{\rm diag}}

\def\la{\lambda}
\def\eps{\epsilon}

\def\ba{\begin{array}}
\def\ea{\end{array}}
\def\beq{\begin{equation}}
\def\eeq{\end{equation}}
\def\be{\begin{equation}}
\def\ee{\end{equation}}

\def\Tr{\mathop{\rm Tr}}

\def\diag{\mathop{\rm diag}}

\def\la{\lambda}
\def\eps{\epsilon}

\newcommand{\bean}{\begin{eqnarray*}}
\newcommand{\eean}{\end{eqnarray*}}

\begin{document}
\thispagestyle{empty} \addtocounter{page}{-1}
   \begin{flushright}
%KIAS-P08nnn \\
%CALT-68-nnnn \\
%{\tt hep-th/yymmnnn}\\
\end{flushright}

\vspace*{1.3cm}
  
 \centerline{ \Large \bf  Comments on Holographic Gravity Dual of } 
\vspace{.3cm} 
\centerline{ \Large \bf  
${\cal N}=6$ Superconformal Chern-Simons Gauge Theory   } 
\vspace*{1.5cm}
\centerline{{\bf Changhyun Ahn }
%, {\bf Kazuo Hosomichi $^{2}$}
%and {\bf Sungjay Lee $^{2}$} 
} 
\vspace*{1.0cm} 
\centerline{\it  
Department of Physics, Kyungpook National University, Taegu
702-701, Korea} 
%\centerline{\it $^{2}$ Korea Institute for 
%Advanced Study, Seoul 130-012, Korea }
\vspace*{0.8cm} 
\centerline{\tt ahn@knu.ac.kr
%, \qquad
%hosomiti@kias.re.kr, \qquad sjlee@kias.re.kr
} 
\vskip2cm

\centerline{\bf Abstract}
\vspace*{0.5cm}

The holographic
nonsupersymmetric renormalization group flows 
in four dimensions are found. 
The mass-deformed ${\cal N}=2, 4$ Chern-Simons matter theories 
can be reproduced from ${\cal N}=1$ Chern-Simons matter theory 
by putting some constraints in the mass terms. 
We construct the geometric superpotential, 
from an eleven dimensional M-theory lift, which 
provides M2-brane probe analysis for the infrared ends of various
supersymmetric or nonsupersymmetric flows.

\baselineskip=18pt
\newpage
\renewcommand{\theequation}
{\arabic{section}\mbox{.}\arabic{equation}}

%%%%%%%%%%%%%%%%%%%%%%%%%%%%%%%%%%%%%%%%%%%%%%%%%%%%%%%%%%%%%%%%%%%%%
%%%%%%%%%%%%%%%%%%%%%%%%%%%%%%%%%%%%%%%%%%%%%%%%%%%%%%%%%%%%%%%%%%%%%%
\section{Introduction}
%%%%%%%%%%%%%%%%%%%%%%%%%%%%%%%%%%%%%%%%%%%%%%%%%%%%%%%%%%%%%%%%%%%%%%
%%%%%%%%%%%%%%%%%%%%%%%%%%%%%%%%%%%%%%%%%%%%%%%%%%%%%%%%%%%%%%%%%%%%%

An explicit construction 
of the renormalization group(RG) flow
between the ultraviolet(UV) fixed point and 
the infrared(IR) fixed point of the three dimensional 
field theory has a close relation to a supergravity kink solution in
four dimensions. 
There exist holographic
RG flow equations connecting ${\cal N}=8$ $SO(8)$ fixed point 
to ${\cal N}=2$ $SU(3) \times U(1)$ fixed point \cite{AP,AW}.
Moreover,  
the other holographic
RG flow equations from ${\cal N}=8$ $SO(8)$ fixed point 
to
${\cal N}=1$ $G_2$
fixed point also exist \cite{AW,AI}.  
The exact solutions to the eleven-dimensional 
bosonic equations corresponding to the $M$-theory lift of these
RG flows are known in \cite{CPW,AI}.

The
three dimensional ${\cal N}=6$ Chern-Simons matter theories
with gauge group $U(N) \times U(N)$ and level $k$ 
have been constructed in \cite{ABJM} and
this theory is described as the low energy limit of $N$ M2-branes at 
${\bf C}^4/{\bf Z}_k$ singularity.
The mass deformed $U(2) \times U(2)$
Chern-Simons gauge theory with level $k=1$ 
or $k=2$ which preserves $SU(3) \times U(1)$ symmetry is studied 
in \cite{Ahn0806n2,BKKS,KKM}. For $G_2$ symmetry case, 
the corresponding mass deformation is described in \cite{Ahn0806n1}.

Besides the above two supersymmetric critical points,
there exist also three nontrivial nonsupersymmetric 
critical points for the scalar
potential:$SO(7)^{+}, SO(7)^{-}$ and $SU(4)^{-}$. 
Although there were some partial attempts in \cite{AW} 
for the finding of RG flow equations
interpolating ${\cal N}=8$ $SO(8)$ fixed point 
to these nonsupersymmetric fixed points, we are going to further 
describe those RG flow 
equations behind those nonsupersymmetric
fixed points.
In order to see the whole structure of the six critical points
including the maximal supersymmetric $SO(8)$ case, we need to use the
$SU(3)$-invariant
sectors(i.e., the $SU(3)$ is a common subgroup of above six invariant
groups) 
of gauged ${\cal N}=8$ supergravity in four dimensions. 
Moreover, the eleven-dimensional metric for the whole
$SU(3)$-invariant sector \cite{AI02}, realized as a warped product of an
asymptotically $AdS_4$ space with a squashed and stretched
seven-sphere,  
is crucial for M2-brane probe.

In this paper, 
starting from the first order differential equations, that are the
nonsupersymmetric flow solutions in four dimensional ${\cal N}=8$ gauged
supergravity
interpolating between an exterior $AdS_4$ region with maximal
${\cal N}=8$ supersymmetry
and an interior $AdS_4$ with no supersymmetry,  
we would like to interpret this as the RG flow in Chern-Simons matter theory
broken to the deformed Chern-Simons matter theory 
by the addition of mass terms for the adjoint superfields. 
An exact correspondence 
can be obtained between fields of bulk supergravity in the $AdS_4$ region
in four-dimensions 
and composite operators of the IR field theory in three-dimensions. 
The three dimensional analog of Leigh-Strassler \cite{LS}
RG flow in mass-deformed Chern-Simons matter 
theory in three dimensions is expected.   
We present the results of probing the eleven-dimensional supergravity
solution corresponding to RG flows.

In section 2, we review the $SU(3)$-invariant 
supergravity solutions in four dimensions
in the context of RG flow, describe the various supergravity critical points 
found previously and present the new nonsupersymmetric flow equations. 
 
In section 3, 
the ${\cal N}=2$ $SU(3) \times U(1)$-invariant 
bosonic mass terms in the Lagrangian  can be reproduced  
by putting the constraints for the mass terms to the 
more generic ${\cal N}=1$ $G_2$-invariant bosonic mass terms.  
Moreover, ${\cal N}=4$ $[SU(2) \times U(1)]^2$-invariant 
bosonic deformed Lagrangian can be constructed. 
We also study ${\cal N}=\frac{1}{2}$ and nonsupersymmetric(${\cal N}=0$)
mass-deformed theories.

In section 4, 
the eleven-dimensional geometric 
superpotential which reduces to the usual $AdS_4$ superpotential for the
particular internal coordinates is described for the M2-brane 
analysis of moduli space.
In particular, the IR behaviors at the nonsupersymmetric critical
points are emphasized. 

In section 5, 
the future directions are presented.

%%%%%%%%%%%%%%%%%%%%%%%%%%%%%%%%%%%%%%%%%%%%%%%%%%%%%%%%%%%%%%%%
%%%%%%%%%%%%%%%%%%%%%%%%%%%%%%%%%%%%%%%%%%%%%%%%%%%%%%%%%%%%%%%%
\section{The holographic RG flows for $SU(3)$-invariant sector 
in four dimensions }
%%%%%%%%%%%%%%%%%%%%%%%%%%%%%%%%%%%%%%%%%%%%%%%%%%%%%%%%%%%%%%%%
%%%%%%%%%%%%%%%%%%%%%%%%%%%%%%%%%%%%%%%%%%%%%%%%%%%%%%%%%%%%%%%%

The gauged ${\cal N}=8$ theory contains self-interaction of a single
massless ${\cal N}=8$ supermultiplet of spins 
$(2, \frac{3}{2}, 1, \frac{1}{2}, 0^{+}, 0^{-})$ with local $SO(8)$ and  local
$SU(8)$ invariance. 
In particular, there exists a non-trivial effective potential for the scalars
that is  proportional to the square of the $SO(8)$ gauge coupling $g$.  
The 70 real, physical 
scalars characterized by $(0^{+}, 0^{-})$ of ${\cal N}=8$ supergravity
parametrize the coset space $E_{7(7)}/SU(8)$ since 63 fields 
may be gauged
away by an $SU(8)$ rotation. 
Then they are described by 
an element ${\cal V}(x)$ of the fundamental 56-dimensional representation
of $E_{7(7)}$:
\bea
{\cal V}(x)=
\left(
\begin{array}{cc}
u_{ij}^{\;\;IJ}(x) & v_{ijKL}(x)  \\
v^{klIJ}(x) & u^{kl}_{\;\;KL}(x)
\end{array} 
\right), 
\nonu
\eea
where $SU(8)$ index pairs $[ij], \cdots$ and $SO(8)$ index pairs $[IJ], \cdots$
are antisymmetrized and $u_{ij}^{\;\;IJ}$ and  $v_{ijKL}$ fields
are $28 \times 28$ matrices. 
Any ground state leaving the symmetry unbroken is necessarily
$AdS_4$ space with a cosmological constant proportional to $g^2$. 
Although the full gauged ${\cal N}=8$ Lagrangian is rather complicated,
the scalar and gravity part of the action  is simple.
Let us define $SU(8)$ T-tensor which is  
manifestly antisymmetric in the indices
$[ij]$ and $SU(8)$ covariant: 
\bea
T_l^{\;kij} & = & 
\left(u^{ij}_{\;\;IJ} +v^{ijIJ} \right) \left( u_{lm}^{\;\;\;JK} 
u^{km}_{\;\;\;KI}-v_{lmJK} v^{kmKL} \right).
\nonu
\eea
This occurs naturally by introducing
a local gauge coupling in the theory. Furthermore, other tensors coming from
T-tensor play an important role and they 
appear in the $g$-dependent interaction terms. 
That is, 
$A_1$ tensor is symmetric in $(ij)$
and $A_2$ tensor is antisymmetric in $[ijk]$: 
\bea
A_1^{\;\;ij} & =& 
-\frac{4}{21} T_{m}^{\;\;ijm}, \;\;\; A_{2l}^{\;\;\;ijk}=-\frac{4}{3}
T_{l}^{\;[ijk]}.
\nonu
\eea
The former appears in the variation of the gravitino of the theory
while the latter appears in the variation of 56 Majorana spinor of the theory.
The $SO(8)$ of gauged supergravity acts on ${\bf R}^8$ as the vector
representation and there exists $SU(3)$ subgroup leaving all the
four-forms invariant where there are three self-dual four-forms and
three anti-self-dual four-forms.
Although there exist six scalar fields in ${\cal N}=8$ supergravity
due to the six four-forms, only four of them are parametrized as the
$SU(3)$ singlet space and they live in the submanifold of $E_{7(7)}/SU(8)$
after using the two residual $U(1)$'s in $SO(8)$ 
and putting  two four-forms to be zero.

The $SU(3)$-invariant sector of the scalar manifold
of gauged ${\cal N}=8$ supergravity \cite{dN82} in four dimensions 
has been studied in \cite{AW}.
The critical points of scalar potential have led to $AdS_4$ vacua and 
the $SU(3)$ gauge symmetry in the supergravity side is preserved.
Then the $SU(3)$-invariant scalar potential of gauged ${\cal N}=8$
supergravity in terms of original variables \cite{Warner83}
can be written as  
\bea 
V(\la, \la',\al, \phi)= g^2 \left[ \frac{16}{3} \left|\frac{\partial
z_3}{\partial \la} \right|^2 + 4 \left|\frac{\partial z_3}{\partial
\la'} \right|^2 - 6  \left|z_3\right|^2 \right] 
\label{scalarpotential}
\eea
where the $z_3$ is the one of the eigenvalues of $A_1$ tensor of the theory
\bea 
z_3(\la, \la',\al, \phi) & = & 6e^{i(\alpha + 2\phi)}p^2qr^2t^2
 + 6e^{2i(\alpha + \phi)}pq^2r^2t^2  + p^3(r^4 + e^{4i\phi}t^4)  \nonu \\
& & +
e^{3i\alpha}q^3(r^4 + e^{4i\phi}t^4).
\label{z3}
\eea
Here we introduce various hyperbolic functions
as follows:
\bea 
& & p \equiv \cosh \left(\frac{\la}{2\sqrt{2}}\right), \;\; q \equiv
\sinh\left(\frac{\la}{2\sqrt{2}}\right), 
\;\; r \equiv \cosh\left(\frac{\la'}{2\sqrt{2}}\right), \;\; t
\equiv \sinh\left(\frac{\la'}{2\sqrt{2}}\right). 
\label{pqrt}
\eea 
Although the equation (\ref{scalarpotential}) doesn't contain 
the derivative terms with respect to the fields $\alpha$ and $\phi$,
one can write down the scalar potential 
in terms of ``true'' superpotential $W$, by using the algebraic
relations found in \cite{AW}
between the complex function $z_3$ and its derivatives with respect to
the fields $\la, \la', \alpha$ and $\phi$,
\bea
V(\la, \la',\al, \phi)  =  g^2 \left[  \frac{16}{3} \left(\partial_{\la}
W \right)^2  + \frac{2}{3p^2 q^2}
\left(\partial_{\al}
W \right)^2 + 4 
\left(\partial_{\la'} 
W  \right)^2+ \frac{1}{2r^2 t^2} \left(\partial_{\phi} 
W \right)^2 - 6  W^2 \right]
\nonu
\eea
with superpotential $W$ which is the magnitude of complex function $z_3$ (\ref{z3}): 
\bea
W(\la, \la', \al, \phi)   =   |z_3|.
\label{superpotential}
\eea

There exist six critical points of this scalar potential. Three of
them are supersymmetric while the other three are nonsupersymmetric.
The symmetry group has a common $SU(3)$ group.

%%%%%%%%%%%%%%%%%%%%%%%%%%%%%%%%%%%%
$\bullet$ $SO(8)$ critical point
%%%%%%%%%%%%%%%%%%%%%%%%%%%%%%%%%%%

This occurs at $\lambda = 0 =\lambda'$, the cosmological constant is
$\Lambda = -6 g^2$(and $W=1$) and 
the ${\cal N}=8$ supersymmetry is preserved.

%%%%%%%%%%%%%%%%%%%%%%%%%%%%%%%%%%%%%%
$\bullet$ $SO(7)^{+}$ critical point
%%%%%%%%%%%%%%%%%%%%%%%%%%%%%%%%%%%%%

This occurs at $\lambda =
\sqrt{2} \sinh^{-1} (\sqrt{\frac{1}{2}(\frac{3}{\sqrt{5}}-1)})=
\lambda'$ and $\alpha=0=\phi$. There is no supersymmetry and the
cosmological constant is given by $\Lambda = -2 \cdot 5^{\frac{3}{4}}
g^2$(and $W=\frac{3}{2} \cdot 5^{-\frac{1}{8}}$).

%%%%%%%%%%%%%%%%%%%%%%%%%%%%%%%%%%%%
$\bullet$ $SO(7)^{-}$ critical point
%%%%%%%%%%%%%%%%%%%%%%%%%%%%%%%%%%%%

This occurs at $\lambda =
\sqrt{2} \, \sinh^{-1} (\frac{1}{2}) =
\lambda'$ and $\alpha= \frac{\pi}{2} =\phi$. There is no supersymmetry and the
cosmological constant is given by $\Lambda = -\frac{25\sqrt{5}}{8}
g^2$(and $W=\frac{3}{8} \cdot 5^{\frac{3}{4}}$).

%%%%%%%%%%%%%%%%%%%%%%%%%%%%%%%%%
$\bullet$ $G_2$ critical point
%%%%%%%%%%%%%%%%%%%%%%%%%%%%%%%%

There is a critical point at
$\la=\sqrt{2}\,\sinh^{-1}(\sqrt{\frac{2}{5}(\sqrt{3}-1)})=\la'$
and $\al=\cos^{-1}(\frac{1}{2}\sqrt{3-\sqrt{3}})=\phi$
and the cosmological constant is $\Lambda=-\frac{216 
\sqrt{2}}{25 \sqrt{5}}\cdot 3^{\frac{1}{4}} g^2
$(and $W=\sqrt{\frac{36\sqrt{2} \cdot 3^{\frac{1}{4}}}{25\sqrt{5}}}$).
This  has an unbroken ${\cal N}=1$ supersymmetry.

%%%%%%%%%%%%%%%%%%%%%%%%%%%%%%%%%%%%
$\bullet$ $SU(4)^{-}$ critical point
%%%%%%%%%%%%%%%%%%%%%%%%%%%%%%%%%%%%

This occurs at $\lambda = 0,
\lambda'= \sqrt{2}\,\sinh^{-1}(1)$ 
and $\phi= \frac{\pi}{2}$. There is no supersymmetry and the
cosmological constant is given by $\Lambda = -8 g^2$(and $W=\frac{3}{2}$).

%%%%%%%%%%%%%%%%%%%%%%%%%%%%%%%%%%%%%%%%%%%%%
$\bullet$ $SU(3) \times U(1)$ critical point
%%%%%%%%%%%%%%%%%%%%%%%%%%%%%%%%%%%%%%%%%%%%%

Finally, there is a critical point at $\lambda = \sqrt{2} \sinh^{-1}
(\frac{1}{\sqrt{3}}), \, \lambda' = \sqrt{2} \sinh^{-1} (\frac{1}{\sqrt{2}}),
\, \alpha=0$ and $\phi=\frac{\pi}{2}$
and the cosmological constant is $\Lambda=-\frac{9 \sqrt{3}}{2} g^2$(and
$W=
\frac{3^{\frac{3}{4}}}{2}$).
This critical point \cite{NW} has an unbroken ${\cal N}=2$ supersymmetry.

For the supergravity description of the 
nonconformal RG flow from one scale to 
another connecting any two critical points, 
the four-dimensional metric which contains 
three dimensional Poincare invariant metric has the form 
$
ds^2= e^{2A(r)} \eta_{\mu' \nu'} dx^{\mu'} dx^{\nu'} + dr^2$
where $\eta_{\mu' \nu'}=(-,+,+)$
and $r$ is the coordinate transverse to the domain wall.
Then the supersymmetric flow equations with (\ref{superpotential}) and 
(\ref{pqrt})
are 
described as \cite{AW}
\bea  
\frac{d \la}{d r} & = & 
\frac{8\sqrt{2}}{3} \, g \, \partial_{\la} W ,\qquad
\frac{d \la'}{d r}   =   
2 \sqrt{2} \, g \, \pa_{\la'} W, \nonu \\
\frac{d \alpha}{d r} & = & 
\frac{\sqrt{2}}{3p^2 q^2} \, g \, \pa_{\al} W, \qquad
\frac{d \phi}{d r}   =   
\frac{\sqrt{2}}{3r^2 t^2} \, g \, \pa_{\phi} W , \nonu \\ 
\frac{d A}{d r} & = & - \sqrt{2} \, g \, W.
\label{flow}
\eea

When we restrict to the case of $G_2$ symmetry, 
the supersymmetric $G_2$ invariant 
reduced flow equations from $SO(8)$ to $G_2$ \cite{AI} can be
described also. On the other hand,
for the $SU(3) \times U(1)$ symmetry case, the supersymmetric reduced 
flow equations 
from $SO(8)$ to $SU(3) \times U(1)$ \cite{AP} are described similarly.

Although the nonsupersymmetric flow equations from $SO(8)$ to $SO(7)^{+}$ in the context
of $SO(5)$ invariant sector are found in \cite{AW}, 
one can easily rederive them in
the present context: $SU(3)$ invariant sector.
By substituting the domain-wall ansatz metric 
into the Lagrangian of scalar-gravity sector, 
the Euler-Lagrangian equations are given in terms of 
the functional $E[A, \la]$.
Then the 
energy-density per unit area transverse to $r$-direction is given by
\bea
 E[A,\la]  =  \int_{-\infty}^{\infty} dr e^{3A}
\left[- 3 \left( 2 (\pa_r A )^2 + \pa_r^2 A \right) - 
\frac{7}{8} \left( \partial_r \la \right)^2
  -V(\la)\right].
\nonu
\eea
One can rewrite this as the sum of complete squares plus others using
the squaring procedure. One arrives at 
\bea
 E[A,\la]  =  \int_{-\infty}^{\infty} dr e^{3A}
\left[ 3 \left( \pa_r A  + \sqrt{2} g W_{+} \right)^2 - 
\frac{7}{8} \left( \partial_r \la  - \frac{8\sqrt{2}}{7} g \pa_{\la} W_{+} \right)^2
 \right] - 
2\sqrt{2} (e^{3A} W_{+})|_{-\infty}^{\infty}.
\nonu
\eea
Then the functional $E[A, \la]$ is extremized by the 
domain-wall solutions.
The first order differential equations, the gradient nonsupersymmetric
flow(we'll explain this further later) 
from $SO(8)$ to $SO(7)^{+}$, 
are written as
\bea  
\frac{d \la}{d r}  = 
\frac{8\sqrt{2}}{7} \, g \, \partial_{\la} W_{+}, \qquad 
\frac{d A}{d r}  =  - \sqrt{2} \, g \, W_{+}
\label{flow2}
\eea
where the superpotential 
$W_{+}$ can be obtained from the generic one $W$ (\ref{superpotential})
by inserting $\alpha=0=\phi$ and $\la=\la'$. The scalar potential is given by
$V(\la)  =  g^2 \left[  \frac{16}{7} \left(\partial_{\la}
W_{+}\right)^2  - 6  W_{+}^2 \right]$. At the supersymmetric $SO(8)$
fixed point, $\partial_{\la} W_{+}$ vanishes while at the
nonsupersymmetric $SO(7)^{+}$ fixed point, it doesn't vanish.

Similarly, 
the nonsupersymmetric flow from $SO(8)$ to $SO(7)^{-}$
can be expressed as
\bea  
\frac{d \la}{d r}  = 
\frac{8\sqrt{2}}{7} \, g \, \partial_{\la} |W_{-}|, \qquad 
\frac{d A}{d r}  =  - \sqrt{2} \, g \, |W_{-}|
\label{flow3}
\eea
where the complex superpotential 
$W_{-}$ can be obtained from the generic complex function $z_3$ (\ref{z3})
by inserting $\alpha=\frac{\pi}{2}=\phi$ and $\la=\la'$ and 
the scalar potential is given by
$V(\la)  =  g^2 \left[  \frac{16}{7} \left|\partial_{\la}
W_{-}\right|^2  - 6  |W_{-}|^2 \right]$.
 At the supersymmetric $SO(8)$
fixed point, $\partial_{\la} |W_{-}|$ vanishes while at the
nonsupersymmetric $SO(7)^{-}$ fixed point, it doesn't vanish.

Compared with the two supersymmetric flows which have two independent fields, 
the nonsupersymmetric flows (\ref{flow2}) and (\ref{flow3})
have only one independent field $\lambda(r)$ 
in addition to the scale function
$A(r)$. Once we choose either $SO(7)^{+}$ or $SO(7)^{-}$, then 
both $\alpha$ and $\phi$ are fixed automatically and only $\lambda(r)$
is left.  

As far as the energy functional procedure we described
before is concerned,
there is no difference between the supersymmetric flows and
nonsupersymmetric flows.
We need to check whether they are supersymmetric or nonsupersymmetric
flows by using other method.
It is known in \cite{ST} that the domain wall solution for which the
scalar is strictly monotonic is supersymmetric for the superpotential.
The domain wall solutions that are asymptotic to unstable $AdS_4$
vacua(that violate Breitenlohner-Freedman bound) are 
nonsupersymmetric. 

One can easily check that the above first order
differential equations (\ref{flow2}) and (\ref{flow3}) 
satisfy the gravitational and scalar equations
of motion by second order differential equations: 
$4\pa_r^2 A + 6 (\pa_r A)^2 + \frac{7}{4} (\pa_r \la)^2 + 2 V =0$
and $\pa_r^2 \la + 3 (\pa_r A) (\pa_r \la) -\frac{4}{7} \pa_{\la} V =0$.
By differentiating the scalar potential $V$, one obtains 
$\pa_{\la} V = 4 g^2 (\pa_{\la} W) (\frac{8}{7} \pa_{\la}^2 W - 3 W)$
where
$W$ is $W_{+}$ or $|W_{-}|$. 
At the critical point of $V$, one has $\pa_{\la} W =0$ or
$\frac{8}{7} \pa_{\la}^2 W = 3 W$.
The $AdS_4$ vacua arising from the former condition 
$\pa_{\la} W =0$(these are also critical points of $W$) 
are supersymmetric and stable by supersymmetry 
while those arising from 
the latter condition $\frac{8}{7} \pa_{\la}^2 W = 3 W$ are 
nonsupersymmetric. The $SO(7)^{\pm}$ critical points hold for this
latter condition as we mentioned before.

According to the observation from \cite{dN84} in old 80's, the two critical 
points $SO(7)^{\pm}$ are unstable against small fluctuations
corresponding to the ${\bf 27}$ representations of $SO(7)$ because the
Breitenlohner-Freedman bound is violated. See \cite{dN84} for detailed
computations on the stability for other representations of $SO(7)$.
The question how a stable domain wall solution can be asymptotic to an
unstable $AdS_4$ vacuum is answered in \cite{ST} in the general context of 
accumulation point of $\pa_r \la$.

Is there any nonsupersymmetric flow from $SO(8)$ to $SU(4)^{-}$?
In this case, the scalar potential is given by 
$V(\la')= g^2 \left[  8 \left(\partial_{\la'}
W_{-}\right)^2  - 8  W_{-}^2 + 2 \right]$ where $W_{-}$ can be
obtained when we put $\la=0$ and 
$\phi=\frac{\pi}{2}$ into the superpotential $W$.
Then, one cannot make the energy density as the sum of complete square
as well as others due to the last term of $V(\la')$. 

We'll come back these flow equations (\ref{flow}), (\ref{flow2}) or 
(\ref{flow3}) when we describe the behavior of
moduli space around IR fixed points in section 4.

%%%%%%%%%%%%%%%%%%%%%%%%%%%%%%%%%%%%%%%%%%%%%%%%%%%%%%%%%%%%%%%%
%%%%%%%%%%%%%%%%%%%%%%%%%%%%%%%%%%%%%%%%%%%%%%%%%%%%%%%%%%%%%%%%
\section{The (non)supersymmetric 
membrane flows in three dimensions }
%%%%%%%%%%%%%%%%%%%%%%%%%%%%%%%%%%%%%%%%%%%%%%%%%%%%%%%%%%%%%%%%
%%%%%%%%%%%%%%%%%%%%%%%%%%%%%%%%%%%%%%%%%%%%%%%%%%%%%%%%%%%%%%%%

Let us recall that the self-dual and anti
self-dual tensors are given by,
in ${\cal N}=8$ gauged supergravity,
\bea
  Y^{1\;\pm}_{ijkl} &=& \varepsilon_{\pm} \left[ \; (\de^{1234}_{ijkl} \pm 
\de^{5678}_{ijkl})+
 (\de^{1256}_{ijkl} \pm \de^{3478}_{ijkl})+(\de^{3456}_{ijkl}
 \pm \de^{1278}_{ijkl}) \; \right],
\nonu \\
       Y^{2\;\pm}_{ijkl} &=& \varepsilon_{\pm} \left[ -(\de^{1357}_{ijkl}
\pm \de^{2468}_{ijkl})+(\de^{2457}_{ijkl} \pm
\de^{1368}_{ijkl})+(\de^{2367}_{ijkl} \pm \de^{1458}_{ijkl}) +
 (\de^{1467}_{ijkl} \pm \de^{2358}_{ijkl}) \right], 
\label{tensor}
\eea
where $\varepsilon_{+} =1$ and $\varepsilon_{-}=i$ and $+$ gives the
scalars while $-$ gives the pseudo-scalars. The indices $ijkl$ refer
to $SU(8)$ indices but after gauge fixing there is no difference
between $SU(8)$ indices and $SO(8)$ indices.
The $SU(3)$ singlet space that is invariant subspace under a
particular $SU(3)$ subgroup of $SO(8)$ has a linear combination of these
four anti-symmetric tensors with four fields 
$\la, \la', \alpha$ and $\phi$ appeared in previous section.
For the $G_2$ symmetric case, one should turn on all these four
tensors $Y^{1\;\pm}_{ijkl}$ and $Y^{2\;\pm}_{ijkl}$ 
while for the $SU(3) \times U(1)$ symmetric case, one should
turn on only $Y^{1+}_{ijkl}$ and $Y^{2-}_{ijkl}$.
This indicates that one expects the mass deformation in boundary
theory,  
preserving the latter, 
can be obtained from the mass deformation preserving the
former by adding some constraints on the mass parameters. 
We'll illustrate this explicitly. 

Let us start with the fermionic mass terms of BL theory \cite{BL0711}:
\bea
{\cal L}_{f.m.} & = & -\frac{i}{2} h_{ab} \bar{\Psi}^a 
\left(  m_1 \Gamma^{78910}+
  m_2\Gamma^{56910}-
  m_3\Gamma^{5678}+
  m_4\Gamma^{46810} \right. \nonu \\
& + & \left.  m_5\Gamma^{4679} +  m_6\Gamma^{4589} -
  m_7\Gamma^{45710} -  m_8 {\bf 1} \right) \Psi^b.
\label{fermionic}
\eea
The indices in eleven-dimensional Gamma matrices 
correspond to the self-dual tensor (or anti self-dual tensor) 
for the indices $5678, 3478, 3456, 2468, 2457, 2367, 2358$, 
if we shift all the
   indices by adding $2$ to (\ref{tensor}), besides an identity.
The corresponding  
fermionic supersymmetric transformation due to the mass deformation
is given by
\bea
\delta_m \Psi^a  & = & \left(   m_1 \Gamma^{78910}+
  m_2\Gamma^{56910}-
  m_3\Gamma^{5678}+
  m_4\Gamma^{46810} \right. \nonu \\
& + & \left.  m_5\Gamma^{4679} +  m_6\Gamma^{4589} -
  m_7\Gamma^{45710} -  m_8 {\bf 1} \right) X_I^a \Gamma_I \epsilon.
\label{mod1}
\eea

Let us classify the possible mass deformations according to the number
of supersymmetry as follows.

%%%%%%%%%%%%%%%%%%%%%%%%%%%%%%%%%%%%%%%%%%%%
$\bullet$ ${\cal N}=1$ supersymmetry 
%%%%%%%%%%%%%%%%%%%%%%%%%%%%%%%%%%%%%%%%%%%%

As done in \cite{Ahn0806n1}(See also \cite{NR}), 
the $\frac{1}{8}$ BPS condition(the number
of supersymmetry is two) 
requires the following constraints on the $\epsilon$ supersymmetry 
parameter 
\bea
\Gamma^{5678}\epsilon  =  \Gamma^{56910}\epsilon=
\Gamma^{78910}\epsilon = \Gamma^{46810} \epsilon 
=
 -\Gamma^{4679} \epsilon=
-\Gamma^{4589}\epsilon= 
-\Gamma^{45710}\epsilon=
-\epsilon.
\label{n1condition}
\eea 
One can easily check this BPS condition by constructing the
eleven-dimensional Gamma matrices explicitly. 

By taking the equal mass condition
\bea
m_1 =m_2 =m_3=m_4=m_5=m_6=m_7=m_8=m,
\nonu
\eea
the bosonic mass term preserving ${\cal N}=1$ supersymmetry
\bea
{\cal L}_{b.m.} = -\frac{1}{2} h_{ab} X_I^a  (m^2)_{IJ} X_J^b
\label{bosonic}
\eea
has the following result which has only one
nonzero component
\bea
(m^2)_{IJ} = \diag(0,0,0,0,0,0,0,64m^2).
\nonu
\eea
After integrating out the massive scalar field at low energy scale,
the sixth order superpotential arises \cite{Ahn0806n1}.

Motivated by the fact that the two M2-branes theory of BL theory is
equivalent to $U(2) \times U(2)$ Chern-Simons matter theory with level
$k=1$ or $k=2$(there is further enhancement from ${\cal N}=6$ 
to ${\cal N}=8$
supersymmetry), 
the natural question is to ask what happens for Chern-Simons
matter theory when we turn on mass perturbation in the gauged supergravity?
Let us consider the $U(2) \times U(2)$ Chern-Simons matter theory.
The theory has matter multiplet in seven flavors $\Phi_i$ where $i =1,
\cdots, 7$
transforming in the adjoint with 
flavor symmetry $G_2$ under which the matter multiplet
forms a septet ${\bf 7}$ of the ${\cal N}=1$ theory.
Now we turn on the mass perturbation in the UV and flow to the IR.
This maps to turning on certain fields in the $AdS_4$
supergravity.
By integrating out the massive scalar $\Phi_8$ that is a singlet ${\bf
1}$ of $G_2$ with adjoint index, 
this results in the 6-th order superpotential $\Tr (Y^{+ijk8} \Phi_i 
\Phi_j \Phi_k)^2 + \Tr \epsilon_{ijklmnp} Y^{+ijk8}(\Phi_l 
\Phi_m \Phi_n \Phi_p)$ \cite{Ahn0806n1}.   
Thus we have found ${\cal N}=1$ superconformal Chern-Simons matter theories
with global $G_2$ symmetry and the Chern-Simons matter theories with 
$G_2$-invariant superpotential deformation are dual to the holographic
RG flows in \cite{AI02}.
We expect that 
$G_2$-invariant $U(N) \times U(N)$ Chern-Simons matter theory 
for $N > 2$ with $k=1,2$ where there exists an enhancement of
${\cal N}=8$ supersymmetry \cite{ABJM,BKKS}  
is dual to the background of \cite{AI02} with $N$ unit of flux.
It would be interesting to explore this direction further.

%%%%%%%%%%%%%%%%%%%%%%%%%%%%%%%%%%%%%%%%
$\bullet$ ${\cal N}=2$ supersymmetry 
%%%%%%%%%%%%%%%%%%%%%%%%%%%%%%%%%%%%%%% 

We impose the constraints on the $\epsilon$ parameter that satisfies the 
$\frac{1}{4}$ BPS condition(the number of supersymmetries is four):
\bea
\Gamma^{5678}\epsilon & = & \Gamma^{56910}\epsilon=
\Gamma^{78910}\epsilon 
= -\epsilon.
\label{cond}
\eea 
Let us further impose the following conditions
\bea
m_1=m_2=m_3=m_8=0.
\label{masscon}
\eea

Using the supersymmetry variation for $X_I^a$,
$\delta X_I^a = i \bar{\epsilon} \Gamma_I \Psi^a$, and
the supersymmetry variation for $\Psi^a$ by the equation (\ref{mod1})
due to the mass deformation,
the variation for  the bosonic mass term (\ref{bosonic}) plus the
fermionic mass term (\ref{fermionic}) under the constraints (\ref{masscon}) leads to
\bea
\delta {\cal L} & = &  i h_{ab} X_I^a  (m^2)_{IJ} \bar{\Psi}^b \Gamma_J \epsilon  
- i h_{ab} \bar{\Psi}^a  \left(   
  m_4\Gamma^{46810}  
 +     m_5\Gamma^{4679} +  m_6\Gamma^{4589} -
  m_7\Gamma^{45710}  \right)^2
X_I^b \Gamma_I \epsilon. 
\nonu
\eea
Then the bosonic mass term $(m^2)_{IJ} \Gamma_J$
should take the form
\bea
(m^2)_{IJ}\Gamma^J & \rightarrow &
 (m_4 -m_5 -m_6 +m_7)^2 (\Gamma_3 +\Gamma_4)
+ (m_4 -m_5 +m_6 -m_7)^2 ( \Gamma_5 +\Gamma_6) \nonu \\
 &+ & (m_4 +m_5 -m_6 -m_7 )^2 ( \Gamma_7 +\Gamma_8)
 +  (m_4 +m_5 +m_6 +m_7)^2 (\Gamma_9 + \Gamma_{10})
\label{massdia}
\eea
where (\ref{cond}) are used.
When all the mass parameters are equal 
$
m_4=m_5=m_6=m_7=m$,
then the diagonal bosonic mass term in (\ref{massdia}) has nonzero
component only for $99$ and $1010$ and other components($33, 44, 55,
66, 77, 88$) are vanishing. The
degeneracy $2$ is related to the ${\cal N}=2$ supersymmetry.  
Then one obtains the bosonic mass term which 
appears in (\ref{bosonic})
\bea
(m^2)_{IJ} = \diag(0,0,0,0,0,0,16m^2,16m^2).
\nonu
\eea
After integrating out the massive scalar field at low energy scale,
the sixth order superpotential occurs \cite{Ahn0806n2}.

The mass deformed BL theory with two M2-branes 
is equivalent to the mass deformed $U(2) \times U(2)$
Chern-Simons gauge theory of \cite{ABJM} with level $k=1$ or $k=2$.
The theory has matter multiplet in three flavors $\Phi_i$ where $i=1,2,3$
transforming in the adjoint with flavor symmetry $SU(3)_I$. 
The $SO(8)_R$ symmetry of the 
${\cal N}=8$ gauge theory is broken to $SU(3)_I \times U(1)_R$
where the former is a flavor symmetry under which the matter multiplet
forms a triplet and the latter is the R-symmetry of the ${\cal N}=2$ theory.
Therefore, we turn on the mass perturbation in the UV and flow to the
IR.
This maps to turning on certain scalar fields in the $AdS_4$
supergravity.
We can integrate out the massive scalar $\Phi_4$ that is a singlet
${\bf 1}$ of $SU(3)_I$ with adjoint index 
at a low enough scale
and this results in the 6-th order superpotential $\Tr 
(f_{abc} f^{ABCD} \Phi_A^a 
\Phi_B^b \Phi_C^c)^2$ \cite{Ahn0806n2}.
The $SU(3)_I \times U(1)_R$-invariant 
$U(N) \times U(N)$ Chern-Simons matter theory 
for $N > 2$ with $k=1,2$ is an open problem.

In this way, the ${\cal N}=2$ superconformal Chern-Simons matter
theory by adding mass terms 
can be constructed from ${\cal N}=1$ superconformal
Chern-Simons matter theory by relaxing the constraints on 
some of the mass terms. 
See also the recent work \cite{BHPW} where
the ${\cal N}=1$ supersymmetric RG flow from $G_2$ symmetric point to 
the $SU(3)_I \times U(1)_R$ symmetric point in BL theory.
Now let us continue to study more
supersymmetric case.

%%%%%%%%%%%%%%%%%%%%%%%%%%%%%%%%%%%%%%%%%%%
$\bullet$ ${\cal N}=4$ supersymmetry 
%%%%%%%%%%%%%%%%%%%%%%%%%%%%%%%%%%%%%%%%%%%

We impose the constraint on the $\epsilon$ parameter that satisfies the 
$\frac{1}{2}$ BPS condition(the number of supersymmetries is eight) \cite{HLL}:
$
\Gamma^{78910}\epsilon = -\epsilon$. 
We impose the additional constraints on $m_4$ and $m_5$ as well as
previous ones (\ref{masscon}):
\bea
m_1=m_2=m_3=m_8=0, \qquad \mbox{and} \qquad m_4=m_5=0.
\nonu
\eea
The variation for  the bosonic mass term  plus the
fermionic mass term  leads to
\bea
\delta {\cal L}  =   i h_{ab} X_I^a  (m^2)_{IJ} \bar{\Psi}^b \Gamma_J \epsilon  
- i h_{ab} \bar{\Psi}^a  \left(   m_6\Gamma^{4589} -
  m_7\Gamma^{45710}  \right)^2
X_I^b \Gamma_I \epsilon. 
\nonu
\eea
Then the bosonic mass term $(m^2)_{IJ} \Gamma_J$
should take the form
\bea
(m^2)_{IJ}\Gamma^J \rightarrow
 (m_6  -m_7)^2 (\Gamma_3 +\Gamma_4 +\Gamma_5 +\Gamma_6)
+ (m_6  +m_7)^2 ( \Gamma_7 +\Gamma_8 + \Gamma_9 +\Gamma_{10}). 
\nonu 
\eea

When the two mass parameters are equal
$
m_6=m_7=m$,
then the diagonal bosonic mass term has nonzero
component only for ($77,88,99$ and $1010$) and other components($33, 44, 55,
66$) are vanishing. The
degeneracy $4$ is related to the ${\cal N}=4$ supersymmetry.  
Then one obtains the bosonic mass term which 
appears in (\ref{bosonic})
\bea
(m^2)_{IJ} = \diag(0,0,0,0,4m^2,4m^2,4m^2,4m^2).
\nonu
\eea

In this way, the ${\cal N}=4$ superconformal Chern-Simons matter
theory by adding mass terms 
can be constructed from ${\cal N}=1$ (or ${\cal N}=2$) superconformal
Chern-Simons matter theory by relaxing the constraints on 
some of the mass terms.
According to the classification for the critical points in previous
section,
there is no ${\cal N}=4$ supersymmetric critical point.
So this theory might be dual to the ${\cal N}=4$ $[SU(2) \times
U(1)]^2$ invariant gauged supergravity
found in \cite{GW}.  
Or one should understand the $SO(3)$ (or $SU(2)$) 
invariant sector of gauged ${\cal N}=8$ supergravity.
We expect 
the full
superpotential $ \frac{M}{2}  \Tr (\Phi_3)^2 +
\frac{M}{2}  \Tr (\Phi_{\hat{4}})^2+ 
 \epsilon^{AB} \epsilon^{\hat{C}\hat{D}} \Tr (\Phi_A \Phi_B
\Phi_{\hat{C}} \Phi_{\hat{D}})$, where $\Phi_A (A =1,3)$ is ${\bf 2}$
representation of one $SU(2)$  and $\Phi_{\hat{C}} (\hat{C}=\hat{2},
\hat{4})$ is ${\bf 2}$ representation of the other $SU(2)$, 
in terms of ${\cal N}=2$ superfields. See also recent work on ${\cal
N}=4$ superfield formalism \cite{CDS}.

Based on the the analysis for the different sector of 
gauged ${\cal N}=8$ supergravity theory found recently, 
let us describe the $U(2) \times U(2)$ Chern-Simons matter theory.
From the superpotential \cite{KKM} of $U(2) \times U(2)$ Chern-Simons
matter theory, the quadratic mass deformations are
added as follows with the notation of \cite{AW09}: 
\bea 
\frac{T^{-4}}{4!} 
 \epsilon_{ABCD} \Tr {\cal Z}^A {\cal Z}^{\ddagger B}
{\cal Z}^C {\cal Z}^{\ddagger D} -2 m^2 T^{-2}  \Tr {\cal Z}^3 {\cal
  Z}^{\ddagger 3} -
2 m^2 T^{-2} \Tr {\cal Z}^4 {\cal Z}^{\ddagger 4},
\nonu
\eea
where ${\cal Z}^A $ is also an ${\cal N}=2$ chiral 
superfield with $SU(4)$ index $A=1, 2, \cdots, 4$, an operation
$\ddagger$ is defined by ${\cal Z}^{\ddagger A} \equiv - i \sigma_2
({\cal Z}^A)^T i \sigma_2$ and 
the $T^2$ is 
a monopole operator.  
The independent two $SU(2)$ transformations, acting on 
the first two components of a complex vector made by a linear
combination of two components of $SO(8)$ vector and on 
the last two components of a complex vector respectively
and  the overall phase rotation gives 
a manifest $[SU(2) \times SU(2)]_R \times U(1)$ symmetry. 
When the undeformed superpotential above is written as 
$\frac{1}{4} \epsilon_{AC} \epsilon^{BD} \Tr {\cal Z}^A {\cal W}_B
{\cal Z}^C {\cal W}_D$,  the $[SU(2) \times SU(2)]_R \times U(1)$ 
symmetry is manifest.
The relation between the deformed Lagrangian from BL
theory
and ${\cal N}=2$ superspace description is evident if we integrate the
superpotential over the fermionic coordinates. 
After the integration over the superspace explicitly, the quartic
terms arise as well as the mass terms for the fermions. 

%%%%%%%%%%%%%%%%%%%%%%%%%%%%%%%%%%%%%%%%%%%%%%%%%%%%%%%%
$\bullet$ ${\cal N}=8$ supersymmetry 
%%%%%%%%%%%%%%%%%%%%%%%%%%%%%%%%%%%%%%%%%%%%%%%%%%%%%%%%

Let us consider the BL theory with $SO(4)$ gauge group and matter fields.
The variation for  the bosonic mass term  
(\ref{bosonic})
plus the
fermionic mass term  
\bea
{\cal L}_{f.m.}  =  -\frac{i}{2} h_{ab} \bar{\Psi}^a 
     m \Gamma^{3456} \Psi^b,
\nonu
\eea
leads to the following variation
\bea
\delta {\cal L}  =   i h_{ab} X_I^a  (m^2)_{IJ} \bar{\Psi}^b \Gamma_J \epsilon  
- i h_{ab} \bar{\Psi}^a  \left(   m \Gamma^{3456}  \right)^2
X_I^b \Gamma_I \epsilon. 
\nonu
\eea
Then the bosonic mass term $(m^2)_{IJ} \Gamma_J$
should take the form
$m^2 \sum_{I=3}^{10} \Gamma_I$. 
Then the diagonal bosonic mass term has nonzero
components for all eight elements.  
Then one obtains the bosonic mass term which 
appears in (\ref{bosonic})
\bea
(m^2)_{IJ} = \diag(m^2, m^2, m^2, m^2, m^2, m^2, m^2, m^2).
\nonu
\eea

Once again we describe the different sector of gauged ${\cal N}=8$ 
supergravity found recently.
From the superpotential of $U(2) \times U(2)$ Chern-Simons
matter theory, the quadratic mass deformations in this case are
added as follows \cite{AW09}: 
\bea 
\frac{T^{-4}}{4!} 
 \epsilon_{ABCD} \Tr {\cal Z}^A {\cal Z}^{\ddagger B}
{\cal Z}^C {\cal Z}^{\ddagger D} 
-\frac{m^2}{2} T^{-2}  \sum_{A=1}^{4} \Tr {\cal Z}^A {\cal
  Z}^{\ddagger A}.
\nonu
\eea      
There exists
a manifest $SU(2) \times SU(2) \times U(1) \times {\bf Z}_2$ symmetry. 
After the integration over the superspace, 
one sees that there are also quartic terms. 
See \cite{AW09} for more details.

Let us describe the case where the supersymmetry is lower than 
${\cal N}=1$ $G_2$ invariant case. 

%%%%%%%%%%%%%%%%%%%%%%%%%%%%%%%%%%%%%%%%%%%%%%%%%%%%%%%%
$\bullet$ ${\cal N}=\frac{1}{2}$ supersymmetry 
%%%%%%%%%%%%%%%%%%%%%%%%%%%%%%%%%%%%%%%%%%%%%%%%%%%%%%%%

We have to introduce more mass
parameters within the structure of (\ref{tensor}). 
Let us consider the following fermionic mass terms 
\bea
{\cal L}_{f.m.} & = & -\frac{i}{2} h_{ab} \bar{\Psi}^a 
\left(  m_1 \Gamma^{78910}+
  m_2\Gamma^{56910}-
  m_3\Gamma^{5678}+
  m_4\Gamma^{46810} + m_5\Gamma^{4679} +  m_6\Gamma^{4589} \right. \nonu \\
& - &   
  m_7\Gamma^{45710} -  m_8 {\bf 1}
 - 
 n_1 \Gamma^{3456}-
  n_2\Gamma^{3478}+
  n_3\Gamma^{34910}-
  n_4\Gamma^{3579} - n_5\Gamma^{35810} \nonu \\
& - & \left.    n_6\Gamma^{36710} +
  n_7\Gamma^{3689}  \right) \Psi^b
\nonu
\eea
where the last seven terms are added newly. Now all the fourteen terms
of
(\ref{tensor}) are present here.
The corresponding fermionic supersymmetry transformation is
\bea
\delta_m \Psi^a  & = & \left(  m_1 \Gamma^{78910}+
  m_2\Gamma^{56910}-
  m_3\Gamma^{5678}+
  m_4\Gamma^{46810} +  m_5\Gamma^{4679} +  m_6\Gamma^{4589} \right. \nonu \\
& - &   
  m_7\Gamma^{45710} -  m_8 {\bf 1}
 - 
 n_1 \Gamma^{3456}-
  n_2\Gamma^{3478}+
  n_3\Gamma^{34910}-
  n_4\Gamma^{3579} \nonu \\
& - & \left.  n_5\Gamma^{35810} -  n_6\Gamma^{36710} +
  n_7\Gamma^{3689}  \right) X_I^a \Gamma_I \epsilon.
\nonu
\eea

We impose the additional seven constraints on the $\epsilon$ parameter
as well as the previous seven condition (\ref{n1condition})
\bea
&& \Gamma^{5678}\epsilon  =  \Gamma^{56910}\epsilon=
\Gamma^{78910}\epsilon = \Gamma^{46810} \epsilon= 
 -\Gamma^{4679} \epsilon=
-\Gamma^{4589}\epsilon= 
-\Gamma^{45710}\epsilon =
\nonu \\
& & 
\Gamma^{34910}
\epsilon
=
\Gamma^{3478}\epsilon= 
\Gamma^{3456}\epsilon =-\Gamma^{35810}
\epsilon
=
-\Gamma^{36710}\epsilon= 
-\Gamma^{3689}\epsilon = \Gamma^{3579}\epsilon
= -\epsilon. 
\label{1cond}
\eea 
Then it turns out this gives rise to $\frac{1}{16}$ BPS 
condition(the number of supersymmetry is one).
The bosonic mass term $(m^2)_{IJ} \Gamma_J$
should take the form
\bea
&& (m_1+m_2-m_3+m_4 -m_5 -m_6 +m_7 -m_8+n_1+n_2-n_3+n_4-n_5-n_6+n_7)^2 \Gamma_3
\nonu \\
&+& (m_1+m_2-m_3-m_4 +m_5 +m_6 -m_7 -m_8+n_1+n_2-n_3-n_4+n_5+n_6-n_7)^2 \Gamma_4
\nonu \\
&+& (m_1-m_2+m_3+m_4 -m_5 +m_6 -m_7 -m_8+n_1-n_2+n_3+n_4-n_5+n_6-n_7)^2 \Gamma_5 \nonu \\
& + & (m_1-m_2+m_3-m_4 +m_5 -m_6 +m_7 -m_8+n_1-n_2+n_3-n_4+n_5-n_6+n_7)^2 \Gamma_6
\nonu \\
 &+ & (m_1-m_2-m_3-m_4 -m_5 +m_6 +m_7 +m_8+n_1-n_2-n_3-n_4-n_5+n_6+n_7)^2 \Gamma_7
\nonu \\
&+ & (m_1-m_2-m_3+m_4 +m_5 -m_6 -m_7 +m_8+n_1-n_2-n_3+n_4+n_5-n_6-n_7)^2 \Gamma_8
\nonu \\
& + & (m_1+m_2+m_3-m_4 -m_5 -m_6 -m_7 +m_8+n_1+n_2+n_3-n_4-n_5-n_6-n_7)^2 \Gamma_9\nonu \\
&+ &
(m_1+m_2+m_3+m_4 +m_5 +m_6 +m_7 +m_8+n_1+n_2+n_3+n_4+n_5+n_6+n_7)^2 \Gamma_{10}.
\nonu
\eea
By taking 
\bea
m_1  &=&  m_2=m_3=m_4=m_5=m_6=m_7=\frac{m_8}{2}=
\nonu \\
n_1 &=& n_2=n_3=n_4=n_5=n_6=n_7= m,
\nonu
\eea
the diagonal bosonic mass term has nonzero
component only for $1010$ and other components are vanishing. 
Then one obtains the bosonic mass term which 
appears in (\ref{bosonic})
\bea
(m^2)_{IJ} = \diag(0,0,0,0,0,0,0, 256m^2).
\nonu
\eea
Note that this form looks like ${\cal N}=1$ supersymmetric case
because there exists only nonzero component for the last element but 
the fermionic mass terms are different from each other.
It would be interesting to construct the gravity dual for this 
${\cal N}=\frac{1}{2}$ mass deformed BL theory.

%%%%%%%%%%%%%%%%%%%%%%%%%%%%%%%%%%%%%%%%%%%
$\bullet$ ${\cal N}=0$ supersymmetry 
%%%%%%%%%%%%%%%%%%%%%%%%%%%%%%%%%%%%%%%%%%%

This can be obtained from (\ref{1cond}) by changing the signs 
of the three constraints in the second line of (\ref{1cond}) as follows:
\bea
&& \Gamma^{5678}\epsilon  =  \Gamma^{56910}\epsilon=
\Gamma^{78910}\epsilon = \Gamma^{46810} \epsilon= 
 -\Gamma^{4679} \epsilon=
-\Gamma^{4589}\epsilon= 
-\Gamma^{45710}\epsilon =
\nonu \\
& & 
\Gamma^{34910}
\epsilon
=
\Gamma^{3478}\epsilon= 
\Gamma^{3456}\epsilon =\Gamma^{35810}
\epsilon
=
\Gamma^{36710}\epsilon= 
\Gamma^{3689}\epsilon = \Gamma^{3579}\epsilon
= -\epsilon. 
\nonu
\eea 
The bosonic mass terms take the form similarly as above
%\bea
%&& (m_1+m_2-m_3+m_4 -m_5 -m_6 +m_7 -m_8+n_1+n_2-n_3+n_4+n_5+n_6-n_7)^2 \Gamma_3
%\nonu \\
%&+& (m_1+m_2-m_3-m_4 +m_5 +m_6 -m_7 -m_8+n_1+n_2-n_3-n_4-n_5-n_6+n_7)^2 \Gamma_4
%\nonu \\
%&+& (m_1-m_2+m_3+m_4 -m_5 +m_6 -m_7 -m_8+n_1-n_2+n_3+n_4+n_5-n_6+n_7)^2 \Gamma_5 \nonu \\
%& + & (m_1-m_2+m_3-m_4 +m_5 -m_6 +m_7 -m_8+n_1-n_2+n_3-n_4-n_5+n_6-n_7)^2 \Gamma_6
%\nonu \\
% &+ & (m_1-m_2-m_3-m_4 -m_5 +m_6 +m_7 +m_8+n_1-n_2-n_3-n_4+n_5-n_6-n_7)^2 \Gamma_7
%\nonu \\
%&+ & (m_1-m_2-m_3+m_4 +m_5 -m_6 -m_7 +m_8+n_1-n_2-n_3+n_4-n_5+n_6+n_7)^2 \Gamma_8
%\nonu \\
%& + & (m_1+m_2+m_3-m_4 -m_5 -m_6 -m_7 +m_8+n_1+n_2+n_3-n_4+n_5+n_6+n_7)^2 \Gamma_9\nonu \\
%&+ &
%(m_1+m_2+m_3+m_4 +m_5 +m_6 +m_7 +m_8+n_1+n_2+n_3+n_4-n_5-n_6-n_7)^2 \Gamma_{10}.
%\nonu
%\eea
and
by taking 
\bea
m_1  &=&  m_2=m_3=m_4=m_5=m_6=m_7=\frac{m_8}{2}=
\nonu \\
n_1 & = & n_2=n_3=n_4=-n_5=-n_6=-n_7= m,
\nonu
\eea
the diagonal bosonic mass term has nonzero
component only for $1010$ and other components are vanishing. 
Then one obtains the bosonic mass term which 
appears in (\ref{bosonic})
\bea
(m^2)_{IJ} = \diag(0,0,0,0,0,0,0, 256m^2).
\label{n0mass}
\eea
We can integrate out the massive scalar $\Phi_8$ with adjoint index 
at a low enough scale
and this results in the sixth order superpotential $\Tr (Y^{\pm ijk8} \Phi_i 
\Phi_j \Phi_k)^2 + \Tr \epsilon_{ijklmnp} 
Y^{\pm ijk8}(\Phi_l \Phi_m \Phi_n \Phi_p)$ with upper sign for
$SO(7)^{+}$
case and lower one for $SO(7)^{-}$ case
in terms of ${\cal N}=1$ superfields. 

For different parametrizations
\bea
m_1 & = & m_2=m_3=m_4=m_5=m_6=m_7=
\nonu \\
-\frac{n_1}{2} & =& -n_2=-n_3=-n_4=n_5=n_6=n_7=
m,  \qquad m_8=0, 
\nonu
\eea
the diagonal bosonic mass term has nonzero
eight components. 
Then one obtains the bosonic mass term which 
appears in (\ref{bosonic})
\bea
(m^2)_{IJ} = \diag(m^2,m^2,m^2,m^2,m^2,m^2,m^2, m^2).
\label{def}
\eea
Then the full superpotential is given by 
the eight mass terms coming from (\ref{def}) 
and quartic terms  from
$ f_{abcd}
Y^{\pm ijkl} \Tr \Phi_i^a \Phi_j^b \Phi_k^c \Phi_l^d$ that are
necessary to the original ${\cal N}=8$ supersymmetry before mass 
deformation.
We turn on the mass perturbation in the UV and flow to the IR.
This maps to turning on certain fields in the $AdS_4$
supergravity where they approach to zero in the UV
and develop a nontrivial profile as a function of
$r$ as one goes to the
IR.

%%%%%%%%%%%%%%%%%%%%%%%%%%%%%%%%%%%%%%%%%%%%%%%%%%%%%%%%%%%%%%%%
%%%%%%%%%%%%%%%%%%%%%%%%%%%%%%%%%%%%%%%%%%%%%%%%%%%%%%%%%%%%%%%%
\section{The potential and metric on the moduli space of an M2-brane probe }
%%%%%%%%%%%%%%%%%%%%%%%%%%%%%%%%%%%%%%%%%%%%%%%%%%%%%%%%%%%%%%%%
%%%%%%%%%%%%%%%%%%%%%%%%%%%%%%%%%%%%%%%%%%%%%%%%%%%%%%%%%%%%%%%%

The eleven-dimensional metric with the warp factor
can be written as \cite{dNW}
\bea
ds_{11}^2 = 
ds_4^2 + ds_7^2 =
\Delta(x,y)^{-1} \,g_{\mu \nu}(x) \,d x^{\mu} d x^{\nu}
+ G_{mn}(x,y) \,dy^m dy^n, 
\label{11metric}
\eea 
where
the warp factor $\Delta(x,y)$ depends on both the four-dimensional
spacetime $x^{\mu}$($\mu=1,2,3,4$) 
and seven-dimensional internal space $y^m$($m=1,2, \cdots, 7$).
The four-dimensional metric which has a three-dimensional 
Poincare invariance
takes the form
$
g_{\mu \nu}(x) \,d x^{\mu} d x^{\nu} = e^{2A(r)} \,\eta_{\mu' \nu'}\,
d x^{\mu'} d x^{\nu'} + dr^2$, 
where $\eta_{\mu' \nu'}=(-,+,+)$ and $r \equiv x^4$ is the coordinate
transverse to the domain wall as in section 2 
and the scale factor $A(r)$ behaves
linearly in $r$ at UV and IR regions. 
The metric formula by \cite{dNW} generates the 7-dimensional metric
$G_{mn}(x,y)$ 
from the four input data of $AdS_4$ vacuum expectation values
for scalar and pseudo-scalar fields $(\la, \la', \al, \phi)$.
We'll use the different parametrizations instead of using 
these fields directly.

Let us introduce the redefined fields as follows:
\bea
a &\equiv& \cosh\!\left(\frac{\lambda}{\sqrt{2}}\right)
+\cos\alpha\,\sinh\!\left(\frac{\lambda}{\sqrt{2}}\right),\nonu \\
b &\equiv& \cosh\!\left(\frac{\lambda}{\sqrt{2}}\right)
-\cos\alpha\,\sinh\!\left(\frac{\lambda}{\sqrt{2}}\right), \nonu \\
c &\equiv& \cosh\!\left(\frac{\lambda'}{\sqrt{2}}\right)
+\cos\phi \,\sinh\!\left(\frac{\lambda'}{\sqrt{2}}\right),\nonu \\
d &\equiv& \cosh\!\left(\frac{\lambda'}{\sqrt{2}}\right)
-\cos\phi \,\sinh\!\left(\frac{\lambda'}{\sqrt{2}}\right).
\label{abcd}
\eea
The seven-dimensional metric turns out to be
\bea
ds_7^2 \equiv G_{mn}\,dy^m dy^n =\sqrt{\Delta}
\left(\frac{a^5 b^5 d^3}{c^3}\right)^{\frac{1}{4}}
L^2 \sum_{i=1}^7 e^{\,i} \otimes e^{\,i},
\label{Gmn}
\eea
where $e^i$ is the local frames in \cite{AI02}, $L$ is a radius of
round
seven-sphere,
and the warp factor is determined as 
\bea
\Delta = \left(\frac{ab}{cd}\right)^{1\over6}\!c^{-1}\xi^{-{4\over3}}.
\label{Delta}
\eea
Here the $SU(3)$ invariant deformed norm on the seven-sphere, from the
deformation matrix, becomes
\bea
\xi^2 = \eta\cos^2\!\mu 
+(\eta_1 \cos^2\!\psi +\eta_2 \sin^2\!\psi)\sin^2\!\mu
\label{xi}
\eea
with deformation parameters
\bea
\eta   =\left(\frac{a^3 b^5 d^3}{c^3} \right)^{\frac{1}{4}},\quad
\eta_1 =\left( \frac{a^7 b  c}{d} \right)^{\frac{1}{4}},\quad
\eta_2 =\left( \frac{a^7 b  d^7}{c^7} \right)^{\frac{1}{4}}.
\label{etas}
\eea
At the $SO(8)$ fixed point where $\eta=1=\eta_1=\eta_2$ because of
$a=b=c=d=1$, 
$\xi^2$
becomes 1. Note that the two coordinates among eight coordinates are
parametrized by 
\bea
X_7 = \sin \mu \cos \psi  \qquad \mbox{and} \qquad  
X_8 = \sin \mu \sin \psi.
\label{78}
\eea
Recall that ${\bf S}^1$ of $U(1)$ Hopf fiber on ${\bf CP}^3$ is
embedded in ${\bf R}^2$ spanned by $X_7$ and $X_8$ and the ${\bf S}^5$
given by Hopf fibration on ${\bf CP}^2$ is embedded in ${\bf R}^6$
spanned by $(X_1, X_2, X_3, X_4, X_5, X_6)$.

Let us describe the ${\cal N}=8$ four-dimensional gauged supergravity.
Now we go to the $SL(8,R)$ basis \cite{KW} and introduce the rotated vielbeins 
\bea
U^{ij}_{\,\,\,\,IJ} & = & u^{ij}_{\,\,\, ab} (\Gamma_{IJ})^{ab}, 
\qquad
V^{ijIJ} = v^{ijab} (\Gamma_{IJ})^{ab}
\nonu \\
U_{ij}^{\,\,\,\,IJ} & = & u_{ij}^{\,\,\, ab} (\Gamma_{IJ})^{ab}, 
\qquad
V_{ijIJ} = v_{ijab} (\Gamma_{IJ})^{ab}
\nonu
\eea
where all indices $i, j$ and $a, b$ run from 1 to 8 and 
correspond to the realization of $E_{7(7)}$ in the $SU(8)$ basis
and $\Gamma_{IJ}$ are the $SO(8)$ generators in \cite{AI}.
We also define the following quantities
\bea
A_{ijIJ} & = & 
\frac{1}{\sqrt{2}} \left( U_{ij}^{\,\,\,\,IJ} + V_{ijIJ}
\right), 
\qquad
B_{ij}^{\,\,\,IJ} =
\frac{1}{\sqrt{2}} \left( U_{ij}^{\,\,\,\,IJ} - V_{ijIJ} \right),
\nonu \\
C^{ij}_{\,\,\,IJ} & = & \frac{1}{\sqrt{2}} \left(U^{ij}_{\,\,\,\,IJ} +
V^{ijIJ}  \right), \qquad 
D^{ijIJ} =\frac{1}{\sqrt{2}} \left(-U^{ij}_{\,\,\,\,IJ} +
V^{ijIJ}  \right).
\nonu
\eea

Then the ``geometric'' T tensor \cite{KW} can be written as
\bea
\widetilde{T}_l^{\,kij} = \frac{1}{168\sqrt{2}} C^{ij}_{\,\,LM} 
\left( A_{lmJK} D^{kmKI} \, \delta^L_{\,I} \, x_M x_J -B_{lm}^{\,\,\,JK} 
C^{km}_{\,\,\,KI} \, \delta^M_{\,J} \, x_L x_I \right)
\nonu
\eea
where we have a relation between $x_I$ and 
$X_I$ that is a coordinate for ${\bf R}^8$:$\sum_{I=1}^{8} (X_I)^2
=1$:
\bea
X_1 & \equiv & \frac{1}{\sqrt{2}} (x_2-x_6), \qquad X_2
\equiv -\frac{1}{\sqrt{2}} (x_3-x_7),
\nonu \\
X_3 & \equiv & \frac{1}{\sqrt{2}} (x_4-x_8), \qquad X_4
\equiv -\frac{1}{\sqrt{2}} (x_1-x_5),
\nonu \\
X_5 & \equiv & \frac{1}{\sqrt{2}} (x_2+x_6), \qquad X_6
\equiv \frac{1}{\sqrt{2}} (x_3+x_7),
\nonu \\
X_7 & \equiv & \frac{1}{\sqrt{2}} (x_4+x_8), \qquad X_8
\equiv \frac{1}{\sqrt{2}} (x_1+x_5).
\nonu 
\eea
From this, the corresponding ``geometric'' $A_1$ tensor is given by
$
\widetilde{A}_1^{\,ij} = \widetilde{T}_m^{\,\,imj}$.

By computing the $88$ component of this $A_1$ tensor, one obtains 
the geometrical superpotential $W_{gs}$ as follows:
\bea
W_{gs}^2 \equiv  |\widetilde{A}_1^{\,88}|^2  & = &  a b^2 c^2 d^2 + 2 a d^2
\left[ b c( a d - b c) -2 (\sqrt{a b -1}-\sqrt{c d -1})^2 \right]
X_7^2 \nonu \\ 
& +&  2 a c^2
\left[ b d( a c - b d) -2 (\sqrt{a b -1}+\sqrt{c d -1})^2 \right] X_8^2
\nonu \\
& + & 2 a \left[ 2 d^2 (\sqrt{a b -1}-\sqrt{c d -1})^2 - 2 c^2 
(\sqrt{a b -1}+\sqrt{c d -1})^2  \right. \nonu \\
& + & \left. 8 a^2 ( 1- c d) + c d (a d - b c)(a
c - b d) + 4 c^2 ( a b + c d -2) \right] X_7^2 X_8^2 \nonu \\
& + & a d^2 \left[ (a d - b c)^2 + 4 (\sqrt{a b -1}-\sqrt{c d -1})^2
\right] X_7^4 \nonu \\
& + &  a c^2 \left[ (a c - b d)^2 + 4 (\sqrt{a b -1}+\sqrt{c d -1})^2
\right] X_8^4.
\label{gsuperpot}
\eea
When the conditions $d=b$ and $c=a$(i.e., $G_2$-invariant sector which
contains $G_2, SO(7)^{\pm}$ critical points) 
are satisfied,
this geometric superpotential does not contain $X_7$ dependence and becomes 
the expression given in \cite{Ahn0806n1}. 
Furthermore, by calculating the $A_2$ tensor obtained from the
geometric T tensor, one arrives at the geometric scalar potential 
\bea
V_{gp}(a,b,c,d,X_7,X_8) & = & 
-g^2 \left( \frac{3}{4} \left| \widetilde{A}_1^{\;ij} \right|^2-\frac{1}{24} \left|
\widetilde{A}^{\;\;i}_{2\;\;jkl}\right|^2 \right) \nonu \\
&=&
8 g^2 a \left[ b \,c\, d + d(a \,d - b \,c) X_7^2  + c( a \,c - b
  \,d) 
X_8^2 \right]^2 
\label{geopot}
\eea
where the constraint equation $\sum_{I=1}^{8} (X_I)^2
=1$ is used several times for obtaining this simple form.
Moreover, the expression $|\widetilde{A}_1^{\,77} + \widetilde{A}_1^{\,88}|^2$
with the conditions $
a  = \frac{1}{b}, c = d$ (i.e., $SU(3) \times U(1)$ critical point) 
leads to 
$16 \widetilde{W}^2$ where $\widetilde{W}$ is 
a geometric superpotential found in \cite{CPW}.
When the conditions $d=b$ and $c=a$(i.e., $G_2$-invariant sector) 
are satisfied,
this geometric scalar potential becomes 
the expression given in \cite{Ahn0806n1}. It is obvious that 
the coefficient of $X_7$ vanishes in (\ref{geopot}) at $G_2$-invariant
sector condition.

When the particular conditions for ${\bf R}^2$ are satisfied
\bea
X_7 = X_8 =\frac{1}{2\sqrt{2}}, \qquad \mbox{or} \qquad
\sin \mu =\frac{1}{2}, \;\;  \mbox{and} \;\; \sin\psi =\frac{1}{\sqrt{2}},
\nonu
\eea 
the geometric superpotential leads to the superpotential $W$(that is, 
$W_{gs} = W$)
where the superpotential $W$ introduced in (\ref{superpotential}) 
can be rewritten, using the relations (\ref{abcd}),  in terms of redefined fields
\bea
W^2 & = & \frac{a}{64} \left[
12 a b(-2+c d)(c^2+d^2) + a^2(16+c^4- 16 c d + 2c^2 d^2 +d^4)
\right. \nonu \\
&-&  12 (2c^3 d + 2 c d^3 - 4 d^2(1+ \sqrt{(ab -1)(cd-1)}) \nonu
  \\ &+& \left. c^2
(-4 -3 b^2 d^2 + 4 \sqrt{(ab -1)(cd-1)}) \right].
\label{Wabcd}
\eea

In terms of $(a, b, c, d)$, the flow equations (\ref{flow}) read in
symmetric form
\bea
\pa_r a & = & \frac{8}{3L} \left[ a^2 \pa_a W + ( a b-2) \pa_b W \right],
\nonu \\
\pa_r b & = & \frac{8}{3L} \left[ (a b-2) \pa_a W + b^2 \pa_b W \right],
\nonu \\
\pa_r c & = & \frac{2}{L} \left[ c^2 \pa_c W + ( c d -2) \pa_d W \right],
\nonu \\
\pa_r d & = & \frac{2}{L} \left[ (c d-2) \pa_c W + d^2 \pa_d W \right],
\nonu \\
\pa_r A & = & -\frac{2}{L} W,
\label{flow1}
\eea
where the superpotential is the same as $AdS_4$ superpotential 
(\ref{superpotential}) but now is given by (\ref{Wabcd}).
Note that the derivatives of $W$ with respect to $a, b, c$ and $d$ do
not vanish at the critical points but the $r$-derivatives of
$a, b, c$ and $d$ do vanish at the critical point because they have
the factors $\frac{d \la}{ d r}, \frac{d \la'}{d r}, \frac{d \alpha}{d
r}$ and $\frac{d \phi}{d r}$ by chain rule.
In other words, for example, the full expression in the first equation
of (\ref{flow1}),
$a^2 \pa_a W + ( a b-2) \pa_b W$, at the critical point vanishes 
even though each term $\pa_a W$ and $\pa_b W$ does not vanish at the
critical point.

The action for the M2-brane probe has two pieces, 
DBI term and WZ term, and it contains 
the pull back metric and three-form onto
the
M2-brane.
We consider a probe that is parallel to the source M2-branes 
and it is traveling at a small velocity transverse to its world
volume.
This leads to a potential and a kinetic term for the M2-brane 
probe. If the potential vanishes then the kinetic term provides us
with a metric on the corresponding moduli space.  
The potential seen by the M2-brane probe \cite{JLP00,KW,JLP01} 
has a factor
\bea
e^{3A} (\Delta^{-\frac{3}{2}} -W_{gs}) & = & e^{3A}
\sqrt{a} \,b \,c \,d\left[ 1+ \left(\frac{a \,c}{b
    \,d} -1\right) X_7^2 + \left(\frac{a \,d}{b \,c}-1\right) X_8^2\right] \nonu \\
&\times & \left( 1- \sqrt{1-\frac{\left[ 1+ (\frac{a \,c}{b
    \,d} -1) X_7^2 + (\frac{a \,d}{b \,c}-1) X_8^2 \right]^2
-\frac{W_{gs}^2}{a \,b^2 \,c^2 \,d^2 }}{\left[1+ (\frac{a \,c}{b
    \,d} -1) X_7^2 + (\frac{a \,d}{b \,c}-1) X_8^2\right]^2}}\right)
\label{potential}
\eea
where we use (\ref{Delta}) and (\ref{gsuperpot}) with the deformed norm 
$\xi^2 =\eta \left[ 1+ \left(\frac{a \,c}{b
    \,d} -1\right) X_7^2 + \left(\frac{a \,d}{b \,c}-1\right)
X_8^2\right]$
from (\ref{xi}) and (\ref{etas}).
The moduli spaces of the brane probe are given by the loci
where the potential vanishes. One sees that 
this potential (\ref{potential}) vanishes at $X_7=0=X_8$. 
By realizing that 
$
X_7 = \sin \mu \cos \psi$ and $ X_8 = \sin \mu \sin
\psi$
appeared in (\ref{78}), 
this leads to $\mu=0$.
On this subspace,  the six-dimensional moduli space
from (\ref{Gmn}), by multiplying the factor $e^A \Delta^{-1/2}$ into
$G_{mn}$, 
is given by
\bea
ds^2|_{\rm{moduli}} = 
\sqrt{a}\, \eta\, 
L^2 \,e^A\left( \sum_{i=1}^4 e^{\,i} \otimes e^{\,i} +  e^{\,7}
  \otimes e^{\,7}
\right)+ \left(
e^A \,\sqrt{a} \, b\, c\,
d \right) dr^2
\label{metric1}
\eea
where the reduced frames are given by
\bea
e^1 &=& \frac{1}{\sqrt{\eta}}  \,d\theta,\qquad
e^2 =   \frac{1}{\sqrt{\eta}}
  \,{\textstyle \frac{1}{2}}
\sin\theta \,\sigma_1,\nonu \\
e^3 &=&   \frac{1}{\sqrt{\eta}}
  \,{\textstyle \frac{1}{2}}
\sin\theta \,\sigma_2,\qquad
e^4 =   \frac{1}{\sqrt{\eta}}
  \,{\textstyle \frac{1}{2}}
\sin\theta \cos\theta \,\sigma_3,\nonu \\
e^7 &=&   
\sqrt{\frac{c \,d}{\eta\, a \,b}}  
\left[d(\phi+\psi)+\frac{1}{2}\sin^2\!\theta\,\sigma_3\right]
\nonu
\eea
and we used the relations  (\ref{Delta}) and (\ref{xi}):
$\Delta = \left(\frac{ab}{cd}\right)^{1\over6}\!c^{-1}\eta^{-{2\over3}}$
and $\xi^2 = \eta$.
The behavior of (\ref{metric1}) can be verified by a numerical study
of the flow equations for the functions $(a(r), b(r), c(r), d(r), A(r))$.
This metric on the moduli space is for $SU(3)$ invariant sector 
containing six critical points.
We have movement on the (squashed) ${\bf S}^5$ with coordinates 
$(\theta,\alpha_1,\alpha_2,\alpha_3,\phi)$ with three Euler angles
$\alpha_i$ 
and the radial direction $r$.

As we approach the IR critical point, a new radial coordinate 
$u \simeq e^{\frac{1}{2} A(r)}$ is introduced.
Moreover, using the flow equations (\ref{flow}), (\ref{flow2}) or (\ref{flow3}),
$\pa_r A=\frac{2}{L} W$ with $g\equiv \frac{\sqrt{2}}{L}$,
one can express the derivative with respect to $r$ in terms of 
$u$. That is, $du^2 = \frac{W^2 u^2}{L^2} dr^2$.   
By substituting this into (\ref{metric1}),  
the six-dimensional moduli space transverse to M2-branes can be
written as
\bea
ds^2|_{\rm{moduli}} =  \frac{L^2 \,\sqrt{a} \, b\, c\,
d }{W^2} \left( du^2 + 
\frac{W^2}{b\, c\, d\, } \, u^2\, ds^2_{FS(2)} + \frac{W^2}{a\, b^2 } \, u^2\,  
\left[d(\phi+\psi)+
\frac{1}{2}\sin^2\!\theta\,\sigma_3\right]^2  \right)
\label{finalmetric}
\eea 
where the Fubini-Study metric on ${\bf CP}^2 (\simeq {\bf S}^4$ when
the angle $\theta$ is very small, i.e., $\cos \theta \simeq 1$) 
is given by $ ds^2_{FS(2)} = d \theta^2 + \frac{1}{4} \sin^2 \theta
(\sigma_1^2 + \sigma_2^2 +\cos^2 \theta \sigma_3^2)$ and $\left[d(\phi+\psi)+
\frac{1}{2}\sin^2\!\theta\,\sigma_3\right]$ is the Hopf fiber on it.
The $\eta$ dependence on the reduced frames $e^1, e^2, e^3$ and $e^4$
has been cancelled by the overall factor $\eta$ in (\ref{metric1}).
Therefore, there is no deformation within  $ ds^2_{FS(2)}$.
At the UV end, $SO(8)$ fixed point where $a=1=b=c=d=W$, 
the metric becomes $ds^2|_{\rm{moduli}} \sim du^2 + u^2 d \Omega_5^2$.
Here $d \Omega_5^2$ is the ${\bf S}^5$ metric given by Hopf fibration
on ${\bf CP}^2$ base.
Moreover, for $X_7=0=X_8$, the three-form potential becomes
$A^{(3)} \sim H^{-1} dt \wedge dx^1 \wedge dx^2$ where the function 
$H=e^{-3A} \Delta^{3/2}$. 

At the IR end of nonsupersymmetric 
$SO(7)^{+}$ flow, by inserting the critical values 
\bea
a = 5^{\frac{1}{4}}=c, \qquad b=5^{-\frac{1}{4}}=d, \qquad 
W=\frac{3}{2} \cdot 5^{-\frac{1}{8}} \qquad : \qquad SO(7)^{+} \; \mbox{symmetry}
\nonu
\eea
into (\ref{finalmetric}), one reads off the coefficients of the
second, third terms of (\ref{finalmetric}) ($\frac{W^2}{b\, c\, d\,}$
and $ \frac{W^2}{a\, b^2 }$) as $\frac{9}{4}$ and
$\frac{9}{4}$
respectively. 
The mass spectrum for the $\frac{\sqrt{7}}{2} \la$ around $SO(7)^{+}$ fixed
point can be computed as in \cite{AR99} and it is 
given by $6$.
At the IR end of the flow, $A(r) \sim \frac{3 
\cdot 5^{-\frac{1}{8}}}{L} r$ and 
$u \sim
e^{\frac{ 3 \cdot 5^{-\frac{1}{8}} }{2 L} r}  
\sim e^{\frac{A(r)}{2}}
$ 
above. 
Then the superfield $S$ becomes $S =(\Phi_1^2  +\cdots + \Phi_7^2 )^
{\frac{9}{4}}$ in the boundary theory. Note that the mass-deformed
bosonic terms are characterized by (\ref{n0mass}). The power $\frac{9}{4}$ comes
from the factor in the metric  of the moduli.
From the tensor product between ${\bf 7}$ and ${\bf 7}$
of $SO(7)^{+}$ representation, one gets a singlet ${\bf 1}$. 
For the superfield $S(x, \theta)$, the 
action looks like $\int d^3 x d^2 \theta 
S(x, \theta)$. 
This implies that the highest component field in $\theta$-expansion
has a conformal dimension  $7$ in the IR. 

At the IR end of nonsupersymmetric 
$SO(7)^{-}$ flow, by inserting the critical values 
\bea
a = \frac{\sqrt{5}}{2}=c, \qquad b=\frac{\sqrt{5}}{2}=d, \qquad 
W=\frac{3}{8} \cdot 5^{\frac{3}{4}} \qquad :
\qquad
SO(7)^{-} \; \mbox{symmetry}
\nonu
\eea
into (\ref{finalmetric}), one reads off the coefficients of the
second, third terms of (\ref{finalmetric}) as $\frac{9}{8}$ and
$\frac{9}{8}$
respectively. 
The mass spectrum for the $\frac{\sqrt{7}}{2} \la$ around $SO(7)^{-}$ fixed
point is 
given by $6$ also.
At the IR end of the flow, $A(r) \sim \frac{ \frac{3}{4} 
\cdot 5^{\frac{3}{4}}}{L} r$ and 
$u \sim
e^{\frac{ \frac{3}{8} \cdot 5^{\frac{3}{4}} }{2 L} r}  
\sim e^{\frac{A(r)}{2}}
$ 
above. 
Then the superfield $S$ is given by $S =(\Phi_1^2  +\cdots + \Phi_7^2 )^
{\frac{9}{8}}$ in the boundary theory. 
From the tensor product between ${\bf 7}$ and ${\bf 7}$
of $SO(7)^{-}$ representation, one gets a singlet ${\bf 1}$. 
The highest component field in $\theta$-expansion
has a conformal dimension  $7$ in the IR. 

At the IR end of supersymmetric 
$SU(3) \times U(1)$ flow, by inserting the critical values 
\bea
a = \sqrt{3} = \frac{1}{b}, \qquad c =\sqrt{\frac{3}{2}} = d, \qquad 
W=
\frac{1}{2}\cdot 3^{\frac{3}{4}} \qquad : \qquad SU(3) \times U(1)
\; \mbox{symmetry}
\nonu
\eea
into (\ref{finalmetric}), one reads off the coefficients of the
second, third terms of (\ref{finalmetric}) as $\frac{3}{2}$ and
$\frac{9}{4}=(\frac{3}{2})^2$
respectively \cite{CPW}: stretching factors.  
The form of IR metric is a consequence of the power $\frac{3}{2}$
of $u^2$ which appears in the Kahler potential \cite{JLP01}.
There is a conical singularity at the origin. 
The ${\bf S}^5$'s in constant $r$ or $u$ radial slices are squashed 
by the presence of $\frac{cd}{ab}$(which is equal to $\frac{3}{2}$
in the IR) instead of unity, after taking out the common factor
$\frac{W^2}{b c d} u^2$ inside of the bracket in (\ref{finalmetric}).  
The residual isometry  or the global symmetry of boundary field theory 
provides this deformation.
Note that the values
$a = 1 =\frac{1}{b}$ and $c= \sqrt{2} =d$ lead to the nonsupersymmetric 
$SU(4)^{-}$ critical point. 

Finally, at the IR end of supersymmetric 
$G_2$ flow, by inserting the critical values 
\bea
a = \sqrt{\frac{6\sqrt{3}}{5}}=c,  \qquad b
=\sqrt{\frac{2\sqrt{3}}{5}}=d, \qquad
W=\sqrt{\frac{36\sqrt{2} \cdot 3^{\frac{1}{4}}}{25\sqrt{5}}}
\qquad : \qquad G_2 \; \mbox{symmetry} 
\nonu
\eea
into (\ref{finalmetric}), one reads off the coefficients of the
second, third terms of (\ref{finalmetric}) as $\frac{6}{5}$ and
$\frac{6}{5}$
respectively \cite{Ahn0806n1}. 

Contrary to the ${\cal N}=2$ $SU(3) \times U(1)$ supersymmetric flow, 
for the $SO(7)^{\pm}$ and $G_2$ critical points, 
the squashing parameter $\frac{cd}{ab}$ inside of ${\bf S}^5$
becomes one due to the fact  that $d=b$ and $c=a$.
The metric looks like  
$ds^2|_{\rm{moduli}} \sim du^2 +  \left(\frac{W^2}{b c d}\right) 
u^2 d \Omega_5^2$.
The values $\frac{9}{4}, \frac{9}{8}$ and $\frac{6}{5}$ coming from 
$\frac{W^2}{b c d}$ arise as above.
Inspite of their difference, both sectors, $SU(3) \times U(1)$
invariant sector and $G_2$ invariant sector  share the same 
${\bf CP}^2$ as mentioned before.

%%%%%%%%%%%%%%%%%%%%%%%%%%%%%%%%%%%%%%%%%%%%%%%%%%%%%%%%%%%%%%%%%%%%%%%%%%%%%%%
%%%%%%%%%%%%%%%%%%%%%%%%%%%%%%%%%%%%%%%%%%%%%%%%%%%%%%%%%%%%%%%%%%%%%%%%%%%%%%%%
\section{
Conclusions and outlook }
%%%%%%%%%%%%%%%%%%%%%%%%%%%%%%%%%%%%%%%%%%%%%%%%%%%%%%%%%%%%%%%%%%%%%%%%%%%%%%%%
%%%%%%%%%%%%%%%%%%%%%%%%%%%%%%%%%%%%%%%%%%%%%%%%%%%%%%%%%%%%%%%%%%%%%%%%%%%%%%%%

We have found the two nonsupersymmetric flow equations preserving 
$SO(7)^{\pm}$, presented its dual theories by adding the mass terms and 
analyzed the M2-brane 
analysis of moduli space with
the IR behaviors.  In particular, (\ref{gsuperpot}) and (\ref{finalmetric})
are necessary to perform this analysis.

It is an open problem to find out 
the eleven-dimensional lift of $SU(3)$ invariant sector 
using (\ref{gsuperpot}). The three-form potential looks like $A^{(3)}
=-e^{3A(r)} W_{gs}(r, \mu, \psi) + \cdots$.
We have to solve the eleven-dimensional Einstein-Maxwell 
equations to complete the eleven-dimensional lift of whole  $SU(3)$-invariant 
sector including RG flows. The eleven-dimensional metric is given by 
(\ref{11metric}) where the compact 7-manifold metric $G_{mn}$ and 
the warp factor $\Delta$ are completely determined by (\ref{Gmn}) and 
(\ref{Delta}) in the local frames. 
The geometric parameters 
$a(r)$, $b(r)$, $c(r)$, $d(r)$ depend on the $AdS_4$ radial coordinate $r$ and are 
subject to the RG flow equations (\ref{flow1}) in 4-dimensional gauged 
supergravity. The local frame is useful to achieve this work. 
As performed 
in \cite{CPW}, one easily makes an ansatz for the 3-form gauge field 
by using the local frames. 

Is there any supersymmetric or nonsupersymmetric 
flow from ${\cal N}=1$ $G_2$ to ${\cal N}=2$ 
$SU(3) \times U(1)$? The negativity of some mass terms
around $G_2$ fixed point supports this possibility.  
Moreover, the symmetry breaking $G_2 \rightarrow SU(3)$ occurs
naturally and this is also other evidence for this RG flow.
If there exists such a flow, then it is interesting to study the behavior
of spectral function given in \cite{Mueck}. 

Although the flow equations around $SU(4)^{-}$ critical point do not
exist,
it is still open problem to find the solution for $A(r)$ around IR
region by numerically along the line of \cite{AR99-1}:
the equations of motion for the scalar and metric satisfy second order
differential equations, in general.
See also recent work on this vacuum \cite{ADFGT}. 

What is the gauged supergravity theory corresponding to
${\cal N}=4$ superconformal Chern-Simons matter theory \cite{N4}? 
As mentioned in section 3, due to the symmetry group $SU(2) \times
SU(2)$,
one needs to search for the $SU(2)$-invariant sector of the gauged 
supergravity and some comments on this possibility 
appeared in \cite{AW02}.
What is the gauged supergravity theory corresponding to
${\cal N}=3$ superconformal Chern-Simons matter theory? 
For other supergravity with Freund-Rubin compactification,
some work is given by \cite{Ahn0810} and see also 
recent paper by \cite{BILPSZ}. For ${\cal N}=1$ case, 
the similar construction is given in \cite{Ahn0809}.
With ${\cal N}=3$ supersymmetry, since 
the $R$ symmetry is $SO(3)=SU(2)$, one has to take more singlets 
among seventy scalars rather than four we considered for
$SU(3)$-invariant sector so far.

What happens when we replace ${\bf CP}^2$ inside the seven-dimensional
internal space with ${\bf S}^2 
\times {\bf S}^2$? The eleven-dimensional lift was given in
\cite{CPW}. This can be done by replacing the stretched 
${\bf S}^5$ with a space which is topologically $T^{1,1}$. 
The observation of \cite{CPW} is that although the eleven-dimensional 
solutions are different but they do have common four-dimensional
gauged ${\cal N}=8$ supergravity. It is an open problem to 
find an eleven-dimensional metric containing ${\bf S}^2 \times {\bf
S}^2$ with common $SU(3)$ invariance.

\vspace{.7cm}

%%%%%%%%%%%%%%%%%%%%%%%%%%%%%%%%%%%%%%%%%%%%%%%%%%%%%%%%%%%%%%
%%%%%%%%%%%%%%%%%%%%%%%%%%%%%%%%%%%%%%%%%%%%%%%%%%%%%%%%%%%%%%%
\centerline{\bf Acknowledgments}
%%%%%%%%%%%%%%%%%%%%%%%%%%%%%%%%%%%%%%%%%%%%%%%%%%%%%%%%%%%%%%%
%%%%%%%%%%%%%%%%%%%%%%%%%%%%%%%%%%%%%%%%%%%%%%%%%%%%%%%%%%%%%%%

I would like to thank 
K. Woo for mathematica computations and
%O. Lunin  and D. Kutasov 
for discussions and K. Skenderis for discussions on the 
domain wall solutions.
This work was supported by grant No.
R01-2006-000-10965-0 from the Basic Research Program of the Korea
Science \& Engineering Foundation.  
%I acknowledge warm hospitality of 
%Particle Theory Group, Enrico Fermi Institute 
%at University of Chicago
%where this work was initiated.

%\appendix
%
%\renewcommand{\thesection}{\large \bf \mbox{Appendix~}\Alph{section}}
%\renewcommand{\theequation}{\Alph{section}\mbox{.}\arabic{equation}}
%\section{The Ricci tensor and field equations}

\end{document}